\documentclass[floatfix,pra,twocolumn,showpacs,amsmath,amssymb,letterpaper,groupaddresses,superscriptaddress]{revtex4}
\usepackage{times}
\usepackage{latexsym}
\usepackage{graphicx}
\usepackage{verbatim,times,bbm}
\usepackage{color}
\usepackage{appendix}

\begin{document}

\title{Entanglement from dissipative dynamics into overlapping environments}

\author{Riccardo Mengoni\footnote{email: riccardo.mengoni@studenti.unicam.it}}
\affiliation{School of Science and Technology, University of Camerino, I-62032 Camerino, Italy}

\author{Laleh Memarzadeh\footnote{email: memarzadeh@sharif.edu}}
\affiliation{Department of Physics, Sharif University of Technology, Teheran, Iran}

\author{Stefano Mancini\footnote{email: stefano.mancini@unicam.it}}
\affiliation{School of Science and Technology, University of Camerino, I-62032 Camerino, Italy}
\affiliation{INFN-Sezione di Perugia, I-06123 Perugia, Italy}

\begin{abstract}
We consider two ensembles of qubit dissipating into two overlapping environments, that is with a certain number of qubit in common that dissipate into both environments.
We then study the dynamics of bipartite entanglement between the two ensembles by excluding the common qubit.
To get analytical solutions for an arbitrary number of qubit we consider initial states with a single excitation and
show that the largest amount of entanglement can be created when excitations are initially located among side
(non common) qubit.
Moreover, the stationary entanglement exhibits a monotonic (resp. non-monotonic) scaling versus the number of  common (resp. side) qubit.
\end{abstract}


\maketitle

\section{Introduction}

The fragility of quantum features has imposed to develop strategies to deal with
unwanted environmental (noisy) effects in quantum information processing.
The standard pursued approach relies in `working against environment', i.e. avoid as much as possible such effects.
However, recently it has been put forward the alternative idea of working `with environment'.
In particular, a dissipative approach to quantum information processing
 may lead to forms of cooperation whereby the environment enhances some coherent tasks performed on the system \cite{VWC09}.

This alternative avenue was paved by studies showing  that even without any interaction among subsystems a common dissipative environment is able to induce entanglement \cite{braun02,benatti03,davidovich,MM11,Cirac11}.
Actually, dissipative systems allow for the stabilization of targeted resources which,
 depending on the task at hand, may results as a key advantage over unitary (noiseless) manipulation.
As matter of fact such a dissipatively generated entanglement can persists up to stationary conditions (see \cite{polzik10} for a recent striking experiment with usage of atomic ensembles).
Ref.\cite{MM13} studied inter-qubit entanglement dynamics
by considering an arbitrary number of qubits dissipating into the same environment.

\begin{figure}[t]
  \centering
       \vspace{-1cm}
   \includegraphics[scale=0.24]{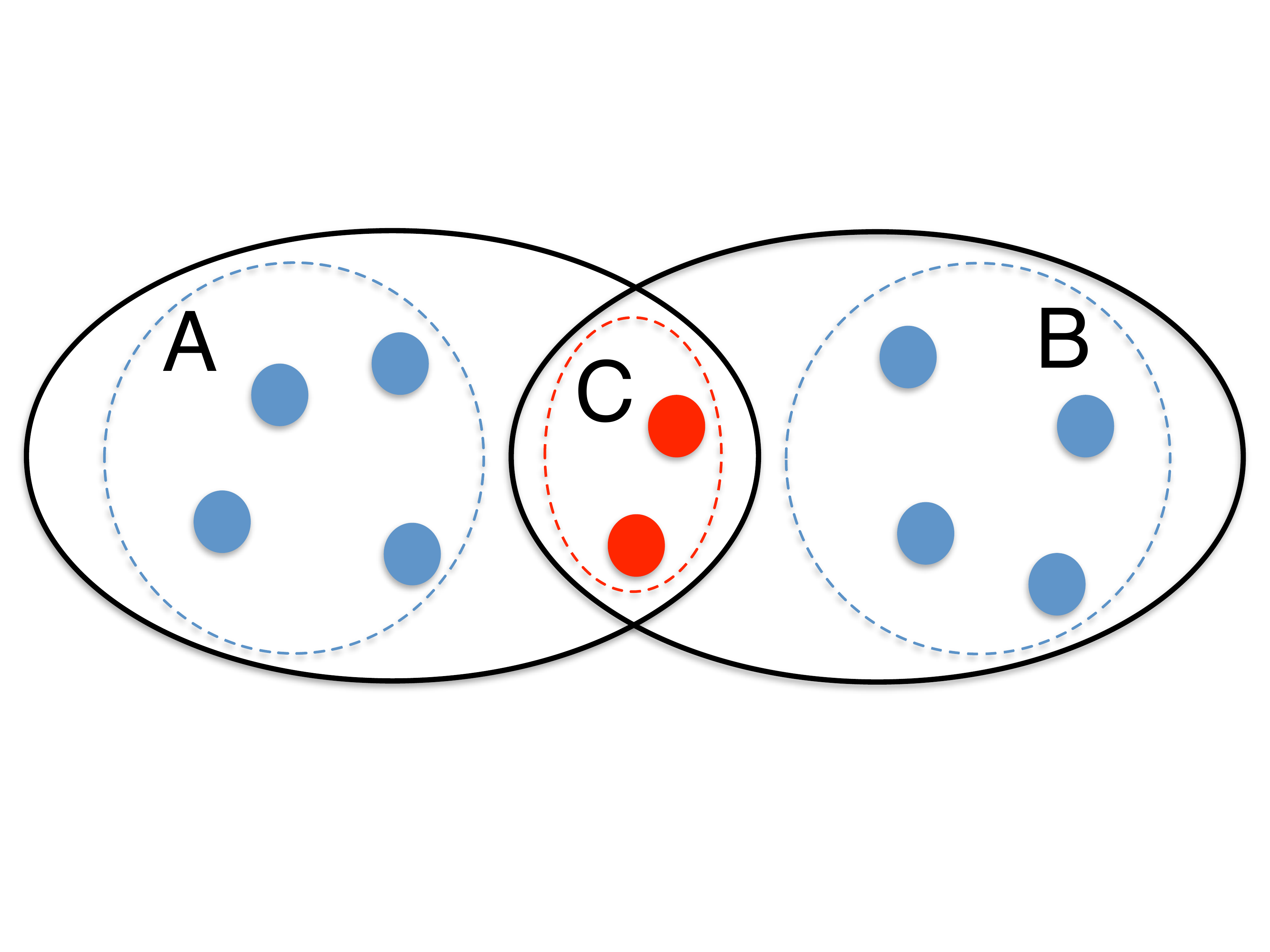}
       \vspace{-1.25cm}
  \caption{(Color online) Pictorial representation of the system under study. An ensemble of qubit $A\cup C$ dissipates into one environment (depicted by solid line); another ensemble of qubit $B\cup C$ dissipates into another environment (depicted by solid line as well). The ensemble of qubit $C$ results dissipating into both environments. }
    \label{scheme}
\end{figure}

Here, along this line, we shall consider a more general scenario in which two ensembles containing arbitrary number of qubits dissipate into overlapping environments (see Fig.\ref{scheme}). It means that a number of qubit will be common to both environments. In this case rather than inter-qubit entanglement it is worth studying the bipartite entanglement between the two ensembles by excluding the common qubit.
To get analytical solutions for an arbitrary number of qubit we consider initial states with a single excitation and
show that the largest amount of entanglement can be created when excitations are initially located among side
(non common) qubit.
Moreover, the stationary entanglement exhibits a monotonic (resp. non-monotonic) scaling versus the number of  common (resp. side) qubit.

The paper is organized as follows. In Section \ref{model} we present the model. Then in Section \ref{ddyn}
we study the dynamics by distinguishing between the cases where the initial excitation is located among
common and side qubit. Going on Section \ref{entanglement} we evaluate the amount of achievable entanglement.
The main results are summarized and discussed in Section \ref{results} and finally conclusions are drawn in Section \ref{conclu}.

%%%%%%%%%%%%%%%%%%%%%%%%%%%%%%%%%%%%%%%%%%
%%%%%%%%%%%%%%%%%%%%%%%%%%%%%%%%%%%%%%%%%%%%%%

\section{The overlapping environments model}\label{model}

Let us consider ensembles $A$, $B$ each containing a number $N$ of qubit and a further ensemble
$C$ containing $n$ qubit \footnote{The case of $A$ containing a different number of qubit with respect to $B$
can be straightforwardly considered, however it leads to much more involved expressions without adding
anything relevant with respect to the situation presented here.}.
Let the qubit belonging to $A\cup C$ dissipate into one environment and those belonging to $B\cup C$ dissipate into another environment (see Fig.\ref{scheme}).
Thus, the ensemble of qubit $C$ results dissipating into both environments.

Given the total number of qubit $N_T=2N+n$, the associated Hilbert space will be $\mathcal{H}\simeq\mathbb{C}^{2\otimes N_T}$ spanned by $\otimes_{i=1}^{N_T}\{|0\rangle _i, |1\rangle _i\}$ with $|0\rangle _i$ and $|1\rangle _i$ representing the ground and excited state of the $i$th qubit.

The purely dissipative dynamics of the entire system will be described by a
Lindbladian master equation \cite{openq} of the following form
\begin{align}\label{Dyn}
\frac{\partial\rho}{\partial t}&=\mathcal{D}[\rho]\notag\\
&\equiv 2\sigma_{AC}\rho\sigma_{AC}^{\dag}-\{\sigma_{AC}^{\dag}\sigma_{AC},\rho\}\notag\\
&+ 2\sigma_{BC}\rho\sigma_{BC}^{\dag}-\{\sigma_{BC}^{\dag}\sigma_{BC},\rho\},
\end{align}
where $\{\;,\;\}$ denotes the anti-commutator,
\begin{equation}
\sigma_{AC}:=\sum_{i\in A\cup C}\sigma_i, \quad \sigma_{BC}:=\sum_{i\in B\cup C}\sigma_i,
\end{equation}
with $\sigma_i:=|0\rangle _i\langle  1|$.
To solve the master equation \eqref{Dyn} we follow the strategy put forward in Ref.\cite{MM13} namely,
starting from the formal solution $\rho(t)=e^{t\mathcal{D}}\rho(0)$ and resorting to the Taylor expansion
of the exponential super operator, we may notice that repeated applications of $\mathcal{D}$ to
$\rho(0)$ will leave the state within a subspace $\mathbb{H}_{\rho(0)}\subset \mathbb{H}$ of the Hilbert space  $\mathbb{H}=\mathcal{H}\otimes\mathcal{H}^*$ (here $\mathcal{H}^*$ stands for the dual of $\mathcal{H}$).
After having identified $\mathbb{H}_{\rho(0)}$, i.e. a set of operators on $\mathcal{H}$ spanning
$\mathbb{H}_{\rho(0)}$, one can write down $\rho(t)$ as linear combination of such operators
with unknown time dependent coefficients.
Then a set of linear differential equations for such coefficients can be derived
 by inserting the expansion back into Eq.\eqref{Dyn}.

The advantage of this procedure is that given a small number of initial excitations $e$
($e\ll N_T$) we have the following inequality {\footnote{This relation corrects Eq.(4) of Ref.\cite{MM13}.}:
\begin{equation}
\label{eqdim}
dim \mathbb{H}_{\rho(0)} \le \left[ \sum_{i=0}^e
\begin{pmatrix}
{N_T}\\{i}\end{pmatrix}
 \right]^2 \ll \left[ 2^{N_T} \right]^2 = dim \mathbb{H}.
\end{equation}

Finally, we also notice from \eqref{Dyn}
that there exist non trivial operators (i.e. not multiple of identity) commuting with the
Lindblad operators $\sigma_{AC}$, $\sigma_{BC}$, hence the stationary solution
will not be unique \cite{Spohn} and we should expect different steady states
depending on $\rho(0)$.

%%%%%%%%%%%%%%%%%%%%%%%%%%%%%%%%%%%%%%%%%%%%%%%%%%%%%%%%%%%%

\section{Dissipative dynamics}\label{ddyn}

Below we confine our attention to the dynamics arising from an initial state containing at most one excitation.
Then given an ensemble $\bullet$ of qubit ($A$, $C$ or $B$ whatever it is), the relevant states will be $|g\rangle^\bullet:=|0\rangle \ldots |0\rangle_i \ldots |0\rangle$
and $|e_i\rangle^\bullet :=|0\rangle \ldots |1\rangle_i \ldots |0\rangle$.
We shall distinguish two cases, one in which such excitation is located in $C$ and the other in which is located in $A$ (or equivalently in $B$).

\subsection{Single excitation initially in $C$}\label{singleC}

Here we assume that in the initial state there exist one excitation among those qubits dissipating energy to both environments (say it is located in the $k$th qubit of ensemble $C$).
We start introducing the following states for the total
$N_T$ qubit system:
\begin{align}
\label{gabck}
|g\rangle &:= |g\rangle^A|g\rangle^C|g\rangle^B,\notag\\
|a\rangle &:=\left(\sum_{i\in A}|{e}_i\rangle^A\right) |g\rangle^C |g\rangle^B,\notag\\
|b\rangle &:=|g\rangle^A|g\rangle^C\left(\sum_{i\in B}|{ e}_i\rangle^B \right),\notag\\
|c\rangle &:= |g\rangle^A\left(\sum_{i\in C \backslash k}|{ e}_i\rangle^C \right) |g\rangle^B,\notag\\
|k\rangle &:= |g\rangle^A |e_k\rangle^C |g\rangle^B.
\end{align}
Actually, $|a\rangle $ (resp. $|b\rangle $) is a uniform superposition of single excitations of qubits belonging to $A$ (resp. $B$) and $|c\rangle $ is a uniform superposition of single excitation of qubits belonging to $C$ excluding the $k$th site.
By applying $\mathcal{D}$ to the states \eqref{gabck} we find the following set of closed relations:
\begin{align}
\label{Daction}
&\mathcal{D}[g\rangle \langle  g|]=0,\cr\cr
&\mathcal{D}[|a\rangle \langle  a|]=2N^2|g\rangle \langle  g|-N(2|a\rangle \langle  a|+\Omega_{ak}+\chi_{ac}),\cr\cr
&\mathcal{D}[|b\rangle \langle  b|]=2N^2|g\rangle \langle  g|-N(2|b\rangle \langle  b|+\Omega_{bk}+\chi_{bc}),\cr\cr
&\mathcal{D}[|c\rangle \langle  c|]=4(n-1)^2|g\rangle \langle  g|\notag\\
&\hspace{1.3cm}-(n-1)(4|c\rangle \langle  c|+2\Omega_{ck}+\chi_{ac}+\chi_{bc}),\cr\cr
&\mathcal{D}[|k\rangle \langle  k|]=4|g\rangle \langle  g|-\Omega_{ak}-\Omega_{bk}-2\Omega_{ck}-4|k\rangle
\langle  k|,\cr\cr
&\mathcal{D}[\Omega_{ak}]=4N|g\rangle \langle  g|-2N|k\rangle \langle  k|-2|a\rangle \langle  a| \notag\\
&\hspace{1.1cm} -(N+2)\Omega_{ak}-N\Omega_{ck}-2\chi_{ac}-\chi_{ab},\cr\cr
&\mathcal{D}[\Omega_{bk}]=4N|g\rangle \langle  g|-2N|k\rangle \langle  k|-2|b\rangle \langle  b| \notag\\
&\hspace{1.1cm}-(N+2)\Omega_{bk}-N\Omega_{ck}-2\chi_{bc}-\chi_{ab},\cr\cr
&\mathcal{D}[\Omega_{ck}]=8(n-1)|g\rangle \langle  g|-4|c\rangle \langle  c| -2n\Omega_{ck}\notag\\
&\hspace{1.1cm}-(n-1)(4|k\rangle \langle  k|+\Omega_{ak}+\Omega_{bk})-\chi_{ac}-\chi_{bc},\cr\cr
&\mathcal{D}[\chi_{ab}]=-2N\chi_{ab}-N(\Omega_{ak}+\chi_{ac}+\Omega_{bk}+\chi_{bc}),\cr\cr
&\mathcal{D}[\chi_{ac}]=-N(2|c\rangle \langle  c|+\Omega_{ck}+\chi_{ac})
+(n-1)\notag\\
&\hspace{1.1cm}\times(4N|g\rangle \langle  g|-2|a\rangle \langle  a|-2\Omega_{ak}-\chi_{ab}-2\chi_{ac}),\cr\cr
&\mathcal{D}[\chi_{bc}]=-N(2|c\rangle \langle  c|+\Omega_{ck}+\chi_{bc})+(n-1)\notag\\
&\hspace{1.1cm}\times(4N|g\rangle \langle  g|-2|b\rangle \langle  b|-2\Omega_{bk}-\chi_{ab}-2\chi_{bc}),\notag\\
\end{align}
where we have defined
\begin{align}
\label{Omchi}
\Omega_{ak}&:=|a\rangle \langle  k|+|k\rangle \langle  a|,
\hskip 4mm 
\chi_{ab}:=|a\rangle \langle  b|+|b\rangle \langle  a|,\notag\\
\Omega_{bk}&:=|b\rangle \langle  k|+|k\rangle \langle  b|,
\hskip 5mm \chi_{bc}:=|b\rangle \langle  c|+|c\rangle \langle  b|,\notag\\
\Omega_{ck}&:=|c\rangle \langle  k|+|k\rangle \langle  c|,
\hskip 5mm 
\chi_{ac}:=|a\rangle \langle  c|+|c\rangle \langle  a|.
\end{align}
The set of closed relations \eqref{Daction} guarantee that the density operator describing the system at arbitrary time $t$ leaves in the space
\begin{align}
\label{Hrho0}
\mathbb{H}_{\rho(0)}=span &\left\{  |g\rangle \langle  g|, |k\rangle \langle  k|, |a\rangle \langle  a|, |b\rangle \langle  b|, |c\rangle \langle  c|, \right.\notag\\
&\left. \Omega_{ak}, \Omega_{bk}, \Omega_{ck}, \chi_{ab}, \chi_{ac}, \chi_{bc} \right\}.
\end{align}
Thus, expanding it as
\begin{align}
\label{exprho}
\rho(t)&=c_0(t)|g\rangle \langle  g|+c_1(t) |k\rangle \langle  k|+c_2(t) |a\rangle \langle  a|\notag\\
&+c_3(t) |b\rangle \langle  b|+c_4(t) |c\rangle \langle  c|+c_5(t) \Omega_{ak}+c_6(t) \Omega_{bk}\notag\\
&+c_7(t) \Omega_{ck}+c_8(t) \chi_{ab}+c_9(t) \chi_{ac}+c_{10}(t) \chi_{bc},
\end{align}
and inserting it back to \eqref{Dyn} yields a set of differential equations for the time dependent coefficients, given in
Appendix \ref{AppA} together with their solutions. 

%%%%%%%%%%%%%%%%%%%%%%%%%%%%%%%%%%%%%%%%%%%%%%%%%%%%%%%%%%

\subsubsection{Tracing out the common qubit}

In order to study entanglement between ensembles $A$ and $B$ we have to first trace out the ensemble $C$.
Taking into account \eqref{gabck}, \eqref{Omchi} and tracing $C$ away from them we get:
\begin{align}
\label{traces}
{\rm Tr}_C (|a\rangle \langle a | )&=
\sum_{i,j\in A}  |e_i\rangle ^A \langle e_j | \otimes |g\rangle ^B \langle g |=: |\tilde{a}\rangle \langle \tilde{a} |,
\notag\\
{\rm Tr}_C (|b\rangle\langle b | )&=  |g\rangle ^A \langle g | \otimes \sum_{i,j\in B}  |e_i\rangle ^B \langle e_j |
=: |\tilde{b}\rangle  \langle \tilde{b}|,
\notag\\
{\rm Tr}_C (|c\rangle  \langle c | )&= (n-1)  |g\rangle ^A \langle g | \otimes |g\rangle ^B \langle g |
\notag\\
&=: (n-1)|\tilde{g}\rangle  \langle \tilde{g}|,
\notag\\
{\rm Tr}_C (|g\rangle  \langle g | )&=|g\rangle ^A \langle  g | \otimes |g\rangle ^B \langle  g |
=|\tilde{g}\rangle  \langle \tilde{g}|,
\notag\\
{\rm Tr}_C (|k\rangle  \langle k | )& = |g\rangle ^A \langle  g | \otimes |g\rangle ^B \langle  g |
= |\tilde{g}\rangle  \langle \tilde{g}|,
\notag\\
{\rm Tr}_C (\chi_{ab})&=|\tilde{a}\rangle \langle  \tilde{b}|+|\tilde{b}\rangle \langle  \tilde{a}|=:\tilde{\chi}_{ab},
\end{align}
while all the other terms in $\{ \Omega_{ak}, \Omega_{bk}, \Omega_{ck},  \chi_{ac}, \chi_{bc}\}$ are zero when the
$ {\rm Tr}_C $ is applied to them.

At the end, thanks to \eqref{traces}, the trace over $C$ of the  density operator \eqref{exprho} gives the following
bipartite state
\begin{equation}
\label{rhoAB}
 \rho_{AB}= \beta(t) |\tilde{g}\rangle  \langle \tilde{g}| +c_2(t)\left(  |\tilde{a}\rangle \langle  \tilde{a}|+ |\tilde{b}\rangle \langle  \tilde{b}| +\tilde{\chi}_{ab}\right),
\end{equation}
where we have taken into account that $ {\rm Tr}(\rho)= c_0+c_1 +2Nc_2+c_4 (n-1)=1$ and defined
\begin{equation}
\label{be}
\beta(t):=1-2N c_2(t).
\end{equation}

%%%%%%%%%%%%%%%%%%%%%%%%%%%%%%%%%%%%%%%%%%%%%%%%%%%%%%%%%%

\subsection{Single excitation initially in $A$}\label{singleA}

We now assume that the initial excitation is in the ensemble $A$ at $k'$th site.
Proceeding like in Sec.\ref{singleC} we introduce, in addition to \eqref{gabck},
the following notation for $N_T$ qubit states:
\begin{align}
\label{gabckprime}
|a'\rangle &:=\left(\sum_{i\in A\backslash k'}|{e}_i\rangle^A \right) |g\rangle^C |g\rangle^B,\notag\\
|c'\rangle &:= |g\rangle^A\left(\sum_{i\in C}|{ e}_i\rangle^C \right) |g\rangle^B,\notag\\
|k'\rangle &:= |e_{k'}\rangle^A |g\rangle^C |g\rangle^B.
\end{align}
Actually, $|a'\rangle $ is a uniform superposition of single excitations of qubits belonging to $A$ excluding the initial excitation at $k'$th and those in $C$. Furtheremore, $|c'\rangle $ is a uniform superposition of single excitation of qubits in $C$.

Using $\mathcal{D}$ of \eqref{Dyn} on \eqref{gabckprime} and \eqref{gabck}
we find the following set of closed relations:
\begin{align}
\label{Dactionprime}
&\mathcal{D}[|k'\rangle \langle  k'|]= 2|g\rangle \langle  g|-\Omega'_{ak}-\Omega'_{ck}-2|k'\rangle \langle  k'|, \cr\cr
&\mathcal{D}[\Omega'_{ak}]= 4(N-1)|g\rangle \langle  g|-2(N-1)|k'\rangle \langle  k'|\notag\\
&\hspace{1.3cm}-2|a'\rangle \langle  a'|-N\Omega'_{ak}-(N-1)\Omega'_{ck}-\chi'_{ac}, \cr\cr
&\mathcal{D}[\Omega'_{bk}]=-(N+1)\Omega'_{bk}-N\Omega'_{ck}-\chi'_{ab}-\chi'_{bc}, \cr\cr
&\mathcal{D}[\Omega'_{ck}]= 4n |g\rangle \langle  g|-n \Omega'_{ak}-(2n+1)\Omega'_{ck}\notag\\
&\hspace{1.3cm}-n\Omega_{bk}'-2n |k'\rangle \langle  k'|-2 |c'\rangle \langle  c'|-\chi'_{ac}, \cr\cr
&\mathcal{D}[\chi'_{ab}]=-(2N-1)\chi'_{ab}-(N-1)(\Omega'_{bk}+\chi'_{bc})-N\chi'_{ac}, \cr\cr
&\mathcal{D}[\chi'_{ac}]=(N-1)(4n |g\rangle \langle  g|-2 |c'\rangle \langle  c'|-\Omega'_{ck})\notag\\
&\hspace{1.3cm}-n (2|a'\rangle \langle  a'|+\Omega'_{ak}+\chi'_{ab})-(N+2n-1)\chi'_{ac}, \cr\cr
&\mathcal{D}[\chi'_{bc}]= n(4N |g\rangle \langle  g|-2|b\rangle \langle  b|-\Omega'_{bk}-\chi'_{ab})\notag\\
&\hspace{1.3cm}-2n \chi'_{bc}-N(2|c'\rangle \langle  c'|+\chi'_{bc}), \cr\cr
&\mathcal{D}[|a'\rangle \langle  a'|]= 2(N-1)^2|g\rangle \langle  g| \notag\\
&\hspace{1.3cm}-(N-1)(2|a'\rangle \langle  a'|+\Omega'_{ak}+\chi'_{ac}), \cr\cr
&\mathcal{D}[|b\rangle \langle  b|]= 2N^2|g\rangle \langle  g|-N(2|b\rangle \langle  b|+\chi'_{bc}), \cr\cr
&\mathcal{D}[|c'\rangle \langle  c'|]= 4n^2|g\rangle \langle  g|-n(4|c'\rangle \langle  c'|+\Omega'_{ck}+\chi'_{ac}+\chi'_{bc}),\notag\\
\end{align}
where, similarly to \eqref{Omchi}, we have defined
\begin{align}
\label{Omchiprime}
\Omega'_{ak}&=|a'\rangle \langle  k'|+|k'\rangle \langle  a'|,
\hskip 4mm 
\chi'_{ab}=|a'\rangle \langle  b|+|b\rangle \langle  a'|, \notag\\
\Omega'_{bk}&=|b\rangle \langle  k'|+|k'\rangle \langle  b|,
\hskip 7mm \chi'_{bc}=|b\rangle \langle  c'|+|c'\rangle \langle  b|, \notag\\
\Omega'_{ck}&=|c'\rangle \langle  k'|+|k'\rangle \langle  c'|,
\hskip 5mm \chi'_{ac}=|a'\rangle \langle  c'|+|c'\rangle \langle  a'|. \notag\\
\end{align}
The set of closed relations in (\ref{Dactionprime}) guarantees that the density matrix describing the system at arbitrary time leaves in the space
\begin{align}
\label{Hrho0prime}
\mathbb{H}_{\rho(0)}=span&\left\{|g\rangle \langle  g|, |k'\rangle \langle  k'|, |a'\rangle \langle  a'|, |b\rangle \langle  b|, |c'\rangle \langle  c'|, \right.\notag\\
&\left.\Omega'_{ak}, \Omega'_{bk}, \Omega'_{ck}, \chi'_{ab}, \chi'_{ac}, \chi'_{bc}\right\}.
\end{align}
Thus, expanding the density matrix as
\begin{align}
\label{rhoexpprime}
\rho(t)&= a_0|g\rangle \langle  g|+a_1 |k'\rangle \langle  k'|+a_2 |a'\rangle \langle  a'|+a_3 |b\rangle \langle  b|\notag \\
 &+a_4 |c'\rangle \langle  c'|+a_5 \Omega'_{ak}+a_6 \Omega'_{bk}+a_7 \Omega'_{ck}+a_8 \chi'_{ab}\notag \\
 &+a_9 \chi'_{ac}+a_{10} \chi'_{bc},
\end{align}
leads (upon insertion into \eqref{Dyn}) to a set of differential equations 
which are reported in Appendix \ref{AppB} together with their solutions. 

%%%%%%%%%%%%%%%%%%%%%%%%%%%%%%%%%%%%%%%%%%%%%%%%%%%%%%%%%%
\subsubsection{Tracing out the common qubit}

In order to study entanglement between ensembles $A$ and $B$ we have to first trace out the ensemble $C$.
Taking into account of  \eqref{gabckprime}, \eqref{Omchiprime} and tracing $C$ away from them we get:
\begin{align}
\label{tracesprime}
{\rm Tr}_C (|a'\rangle \langle a' | )&=
\sum_{i,j\in A\backslash k'}  |e_i\rangle ^A \langle e_j | \otimes |g\rangle ^B \langle g |
=: |\tilde{a'}\rangle \langle \tilde{a'} |,
\notag\\
{\rm Tr}_C (|b\rangle\langle b | )&= |g\rangle ^A \langle g | \otimes \sum_{i,j\in B}  |e_i\rangle ^B \langle e_j |
=: |\tilde{b}\rangle  \langle \tilde{b}|,
\notag\\
{\rm Tr}_C (|c'\rangle  \langle c' | )&= n  |g\rangle ^A \langle g | \otimes|g\rangle ^B \langle g |
=: n|\tilde{g}\rangle  \langle \tilde{g}|,
\notag\\
{\rm Tr}_C (|g\rangle  \langle g | )&=|g\rangle ^A \langle  g | \otimes |g\rangle ^B \langle  g |
=:|\tilde{g}\rangle  \langle \tilde{g}|,
\notag\\
{\rm Tr}_C (|k'\rangle  \langle k' | )& = |e_{k'}\rangle^A \langle  e_{k'} | \otimes |g\rangle ^B \langle  g |
=: |\tilde{k'}\rangle  \langle \tilde{k'}|,
\notag\\
{\rm Tr}_C (\chi'_{ab})&=|\tilde{a'}\rangle \langle  \tilde{b}|+|\tilde{b}\rangle \langle  \tilde{a'}|=:\tilde{\chi'}_{ab},
\notag\\
{\rm Tr}_C (\Omega'_{ak})&=|\tilde{a'}\rangle \langle  \tilde{k'}|+|\tilde{k'}\rangle \langle  \tilde{a'}|=:\tilde{\Omega'}_{ak},
\notag\\
{\rm Tr}_C (\Omega'_{bk})&=|\tilde{b}\rangle \langle  \tilde{k'}|+|\tilde{k'}\rangle \langle  \tilde{b}|=:\tilde{\Omega'}_{bk},
\end{align}
while all the other terms in $\{ \Omega_{ck},  \chi_{ac}, \chi_{bc}\}$ are zero when the
$ {\rm Tr}_C $ is applied to them.

At the end, thanks to \eqref{tracesprime}, the trace over $C$ of the  density operator \eqref{rhoexpprime} gives the following
bipartite state
\begin{align}
\label{rhoABprime}
 \rho_{AB}&={\rm Tr}_C\left(\rho \right)\notag\\
 &=\beta' |\tilde{g}\rangle \langle\tilde{g}|+a_1 |\tilde{k'}\rangle \langle\tilde{k'}|
+a_2  |\tilde{a'}\rangle \langle \tilde{a'}|+ a_3|\tilde{b}\rangle \langle \tilde{b}|\notag\\
&+a_5\tilde{\Omega'}_{ak}+a_6\tilde{\Omega'}_{bk}  +a_8 \tilde{\chi'}_{ab},
\end{align}
where we have used the relation ${\rm Tr}(\rho)= a_0+a_1 +(N-1)a_2+Na_3+n a_4=1 $ and defined
\begin{equation}
\label{beprime}
\beta':=1-a_1-(N-1)a_2-N a_3.
\end{equation}

%%%%%%%%%%%%%%%%%%%%%%%%%%%%%%%%%%%%%%%%%%%

\section{Evaluating the amount of entanglement}\label{entanglement}

To evaluate the amount of entanglement between ensembles $A$ and $B$
we use the negativity introduced in \cite{VW02} and later proved as a valid entanglement monotone \cite{P05}.
Since the negativity is defined using the partial transposition we have to find $ \rho_{AB}^{T_B} $.
Again we distinguish two situations according to Sections \ref{singleC}, \ref{singleA}.

\subsection{Single excitation initially in $C$}

We first derive from \eqref{traces} the following result:
\begin{align}
\label{TB}
\left( |\tilde{a}\rangle  \langle \tilde{b}|\right) ^{T_B}
&=\sum_{i\in A}\sum_{j\in B} |e_i\rangle ^A \langle  g|  \otimes \left( |g\rangle ^B \langle e_j |\right) ^{T_B}\notag\\
& = |\tilde{ab}\rangle  \langle \tilde{g}|,
\end{align}
where we have introduced the state
\begin{equation}
\label{tildeab}
 |\tilde{ab}\rangle :=\sum_{i\in A }\sum_{j\in B } |{e}_i\rangle ^A \otimes |{e}_j\rangle ^B .
 \end{equation}
Then, using \eqref{TB} in \eqref{rhoAB} we get
\begin{align}
\label{rhoPT}
 \rho_{AB}^{T_B}&= \beta |\tilde{g}\rangle  \langle \tilde{g}| \notag\\
& +c_2\left(  |\tilde{a}\rangle \langle  \tilde{a}|+ |\tilde{b}\rangle \langle  \tilde{b}| +|\tilde{g}\rangle  \langle \tilde{ab}|+|\tilde{ab}\rangle  \langle \tilde{g}|\right),
\end{align}
with $\beta$ defined in \eqref{be}.

The negativity is equal, by definition, to the absolute value  of the sum of the negative eigenvalues of
$ \rho_{AB}^{T_B}  $.
In order to find these eigenvalues, recalling the definition of $ |\tilde{g}\rangle ,|\tilde{ab}\rangle ,|\tilde{a}\rangle ,|\tilde{b}\rangle$ given in \eqref{traces}, we can represent $ \rho_{AB}^{T_B}  $ in the basis
$ \left\lbrace |\tilde{g}\rangle ,|{ e}_i\rangle ^A \otimes |{ e}_j\rangle ^B ,|{ e}_i\rangle ^A \otimes |g\rangle ^B,
|g\rangle ^A \otimes |{ e}_j\rangle ^B  \right\rbrace_{i\in A, j\in B}  $ which includes $ 1+N^2+2N  $ vectors.
Then, $ \rho_{AB}^{T_B}  $ takes the following block matrix form
\begin{equation}
\rho_{AB}^{T_B}=\begin{pmatrix}
\tau& 0\\ 0& \omega
\end{pmatrix},
\end{equation}
where $ \tau $ and $ \omega $ are matrices of dimensions $ (1+N^2)\times (1+N^2) $ and $ (2N)\times (2N) $ respectively, made in the following way:
\begin{equation}
\tau:=\begin{pmatrix}
\beta & c_2&\ldots& c_2\\
c_2   & 0    &\ldots& 0     \\
 \vdots    &   & \ddots & \vdots \\
 c_2   & 0    &\ldots& 0
 \end{pmatrix},
\end{equation}
and
\begin{equation}
\omega:=\begin{pmatrix}
\omega^{(1)}& 0\\ 0& \omega^{(1)}
\end{pmatrix},
\end{equation}
with $ \omega^{(1)}$ a $ N \times N $ matrix having all entries equal to $ c_2(t) $, i.e.
\begin{equation}
\omega^{(1)}:=\begin{pmatrix}
c_2&c_2&  \ldots & c_2& c_2\\
c_2&c_2&  \ldots & c_2& c_2\\
 \vdots    &                & \ddots& & \vdots \\
 c_2&c_2&  \ldots & c_2& c_2\\
 c_2&c_2&  \ldots & c_2& c_2
\end{pmatrix}.
\end{equation}
For the  property of block diagonal matrix determinant, the eigenvalues of $ \rho_{AB}^{T_B} $ satisfy the relation
\begin{equation}
\det\left( \tau -\lambda I_{N^2+1}\right)\det\left( \omega^{(1)} -\lambda I_N\right)
\det\left( \omega^{(1)} -\lambda I_N\right)=0,
\end{equation}
($I_N$ denotes the $N\times N$ identity matrix).
 Thus, the non-zero eigenvalues of $ \rho_{AB}^{T_B} $ are those of the two matrices $ \tau $ and $ \omega^{(1)}$, namely
 \begin{align}
 \label{eigenval}
 \lambda (\tau)=& \dfrac{1}{2}\left(   \beta(t)\pm\sqrt{(2Nc_2(t))^2+\beta(t)^2} \right), \notag\\
 \lambda (\omega^{(1)})=& N c_2(t).
 \end{align}
The only negative eigenvalue is the one of $\tau $ with the minus in front of the square root, hence the negativity
results
\begin{equation}
\label{negC}
{\mathcal{N}}(t)=\sqrt{\left(N c_2(t)\right)^2+\left(\frac{1}{2}-N c_2(t) \right)^2}
- \left( \frac{1}{2}-N c_2(t)\right),
\end{equation}
where the relation $ \beta(t)=1-2N c_2(t) $  has been used (see \eqref{be}).

%%%%%%%%%%%%%%%%%%%%%%%%%%%%%%%%%%%%%%%%%%%%%%

\subsection{Single excitation initially in $A$}

We first derive from \eqref{tracesprime} the  following results:
\begin{align}
\label{TBprime}
\left( |\tilde{a'}\rangle  \langle \tilde{b}|\right) ^{T_B} &=\sum_{i\in A \backslash k'}\sum_{j\in B} |e_i\rangle ^A\langle  g|  \otimes \left( |g\rangle ^B \langle e_j |\right) ^{T_B}
\notag\\
& = |\tilde{a'b}\rangle  \langle \tilde{g}|,
\notag\\
\left( |\tilde{k'}\rangle  \langle \tilde{b}|\right) ^{T_B} &=\sum_{j\in B} |e_{k'}\rangle ^A \langle  g|  \otimes
\left( |g\rangle ^B \langle e_j |\right) ^{T_B} = |\tilde{k'b}\rangle  \langle \tilde{g}|,
\notag\\
\end{align}
where we have introduced the states
\begin{equation}
\label{tildeabprime}
\begin{split}
 |\tilde{a'b}\rangle &:=\sum_{i\in A \backslash k' }\sum_{j\in B } |e_i\rangle ^A \otimes |e_j\rangle ^B,\\
  |\tilde{k'b}\rangle&:=\sum_{j\in B} |e_{k'}\rangle^A \otimes |e_j\rangle^B.
  \end{split}
\end{equation}
Finally, using \eqref{TBprime} in \eqref{rhoABprime}, we get
\begin{equation}\begin{split}
\rho_{AB}^{T_B}
&=\beta' |\tilde{g}\rangle \langle\tilde{g}|+a_1 |\tilde{k'}\rangle \langle\tilde{k'}|\\
&+a_2  |\tilde{a'}\rangle\langle \tilde{a'}|+ a_3|\tilde{b}\rangle\langle \tilde{b}|+a_5\tilde{\Omega'}_{ak}\\
&+a_6\left(|\tilde{g}\rangle \langle\tilde{k'b}|+|\tilde{k'b}\rangle \langle\tilde{g}| \right)   +a_8 \left( |\tilde{a'b}\rangle \langle\tilde{g}|+|\tilde{g}\rangle \langle\tilde{a'b}|\right),
\end{split}
\end{equation}
with $\beta'$ defined in \eqref{beprime}.

The negativity is equal, by definition, to the absolute value  of the sum of the negative eigenvalues of
$ \rho_{AB}^{T_B}  $.
In order to find these eigenvalues, recalling the definition of $ |\tilde{g}\rangle ,|\tilde{ab}\rangle ,|\tilde{a}\rangle ,|\tilde{b}\rangle$ given in \eqref{traces}, we can represent $ \rho_{AB}^{T_B}  $ in the basis
$ \left\lbrace |\tilde{g}\rangle ,|e_i\rangle ^A \otimes |e_j\rangle ^B ,|e_i\rangle ^A \otimes |g\rangle ^B,
|g\rangle ^A \otimes |e_j\rangle ^B \right\rbrace_{i\in A, j\in B}  $ which includes $ 1+N^2+2N  $ vectors.
Then, $ \rho_{AB}^{T_B}  $ takes the following block matrix form
\begin{equation}
\rho_{AB}^{T_B}=\begin{pmatrix}
\tau'& 0\\ 0& \omega'\end{pmatrix},
\end{equation}
where $ \tau' $ and $ \omega' $ are matrices of dimensions $ (1+N^2)\times (1+N^2) $ and $ (2N)\times (2N) $ respectively, made in the following way:
\begin{equation}
\tau'=\begin{pmatrix}
\beta' & a_8&\ldots & \ldots & a_8& a_6& \ldots& a_6\\
a_8   & 0     &\ldots & \ldots& 0    & 0     & \ldots & 0\\
 \vdots &   &     \ddots& & & & &\vdots \\
  \vdots &   &  &     \ddots& & & &\vdots \\
 a_8   & 0     &\ldots & \ldots & 0    & 0     & \ldots & 0\\
 a_6   & 0     &\ldots & \ldots & 0    & 0     & \ldots & 0\\
 \vdots    & &  &  &      &     & \ddots &\vdots \\
  a_6   & 0     &\ldots & \ldots &  0    & 0     & \ldots & 0
  \end{pmatrix},
\end{equation}
(there are $N(N-1)$ elements equal to $ a_8 $ and $N$ elements equal to $ a_6 $ in the first row and column ) and
\begin{equation}
\omega'=\begin{pmatrix}
\omega'^{(1)}& 0\\ 0& \omega'^{(2)}
\end{pmatrix},
\end{equation} 
where $ \omega'^{(1)}  $ and $ \omega'^{(2)} $ are  $ N\times N $ matrices made as follows
\begin{equation}
\omega'^{(1)}=\begin{pmatrix}
a_1& a_5& a_5&  ...& a_5& a_5\\ a_5& a_2& a_2&  ...& a_2& a_2\\ a_5& a_2& a_2&  ...& a_2& a_2\\
 \vdots    &                & & \ddots& & \vdots \\a_5& a_2& a_2&  ...& a_2& a_2\\ a_5& a_2& a_2&  ...& a_2& a_2\\  \end{pmatrix},
\end{equation}
and
\begin{equation}
\omega'^{(2)}=\begin{pmatrix}
a_3& a_3&  ...& a_3& a_3\\ a_3& a_3&  ...& a_3& a_3\\
 \vdots    &                & \ddots& & \vdots \\ a_3& a_3& ...& a_3& a_3\\ a_3& a_3& ...& a_3& a_3
 \end{pmatrix}.
\end{equation}
For the  property of block diagonal matrix determinant, the eigenvalues of $ \rho_{AB}^{T_B} $ satisfy the equation
\begin{equation}
\det\left( \tau' -\lambda I_{N^2+1}\right) \det\left( \omega'^{(1)} -\lambda I_N\right)\det\left( \omega'^{(2)} -\lambda I_N\right)=0.
\end{equation}
 Thus the non-zero eigenvalues of $ \rho_{AB}^{T_B} $ are those of the  matrices $ \tau'$, $ \omega'^{(1)} $ and $ \omega'^{(2)}$, namely
 \begin{align}
 \label{eigenval2}
 \lambda (\tau')&= \frac{1}{2}(\beta')\pm \frac{1}{2}\sqrt{(\beta')^2+4 N \left(a_6^2+(N-1)a_8^2\right)} , \notag\\
 \lambda (\omega'^{(1)})&=  \frac{1}{2}\left( a_1+(N-1)a_2\right)\notag\\
 &\pm \sqrt{(a_1-(N-1)a_2)^2+4(N-1)a_5^2 }, \notag\\
 \lambda (\omega'^{(2)}) &= N a_3.
 \end{align}

The only negative eigenvalue is the one of $ \tau' $ with the minus in front of the square root, hence the negativity
results
\begin{equation}
\label{negA}
\begin{split}
&\mathcal{N}(t)=\\ &\frac{1}{2}\sqrt{(a_1+a_2 (N-1)+a_3 N-1)^2+4 N \left(a_6^2+(N-1)a_8^2\right)}\\&-\frac{1}{2}(1-a_1-a_2 N+a_2-a_3 N).
\end{split}
 \end{equation}

%%%%%%%%%%%%%%%%%%%%%%%%%%%%%%%%%%%%%%%%%%%%%%

\section{Summary of results}\label{results}

Let us now comment the main results of this paper which stem form the negativity expressions \eqref{negC} and \eqref{negA}. 
First notice that they coincide only in the case of $N=n=1$.

Then, the negativity \eqref{negC} results monotonically increasing vs time up to a stationary value.
In the limit $t\to\infty$ it becomes
\begin{equation}
\label{negCst}
\begin{split}
\mathcal{N}(\infty)&=\frac{1}{2} \sqrt{\frac{8 N^2}{(2 n+N)^4}-\frac{4 N}{(2
   n+N)^2}+1}\\
   &+\frac{N}{(2 n+N)^2}-\frac{1}{2}.
 \end{split}
\end{equation}
The above quantity monotonically decreases vs $n$.
This can be explained by the fact that the initial single excitation for increasing $n$
tends to persists in the common part rather than being shared by the side parts.
Furthermore, \eqref{negCst} is non monotonic vs $N$.
Actually it has a maximum for $N=2n$.

Coming to the negativity \eqref{negA}, it also results monotonically increasing vs time up to a stationary value.
In the limit $t\to\infty$ it  becomes
\begin{equation}
\label{negAst}
\begin{split}
\mathcal{N}(\infty)&=
\frac{1}{2N^2(N+2n)^2}
\left[N^6+4 N^5 n (n+2)\right.\\
&\left.+16 N^4 n^2
   (n+1)+16 N^3 n^4-8 N^2 n^4
\right]^{1/2}\\
&+\frac{2 n^2-(N+2
   n)^2}{2 N (N+2 n)^2}.
\end{split}
\end{equation}
In contrast to \eqref{negCst}, the negativity \eqref{negAst} monotonically increases vs $n$ and reaches a  saturation value only for $ n\gg N $.
This can be explained by the fact that the initial single excitation for increasing $n$
is more easily shared by the side parts -- due to the increasing common part.
Finally, also \eqref{negAst} is non monotonic vs $N$.
Actually it has a maximum for $N=2$, whatever the value of $n$ is.

The non monotonic behavior of both \eqref{negC} and \eqref{negA} vs $N$ should be ascribed to competing effects 
of side and common qubits.

%%%%%%%%%%%%%%%%%%%%%%%%%%%%%%%%%%%%%%%%%%%%%%

\section{Conclusion}\label{conclu}

In conclusion, we have studied entanglement arising in the dynamics of two qubit ensembles 
dissipating into overlapping environments. That is, having a number of qubit in common to both environments. 
We have computed the bipartite entanglement between the two ensembles by excluding such common qubit 
and asuming a single initial excitation.

Our study shows that the dynamics of the entanglement crucially depends on the initial condition, especially
on whether the single initial excitation is in the common qubit $C$ or in the side qubit $A$ or $B$.
Furthermore, the amount of entanglement that can be dissipatively created depends on the number $n$ of common qubit. We have also characterized the stationary properties of such entanglement.

The studied model might be of interest for several physical systems.
Simulation of dissipative dynamics of small ensembles of qubits has been already engineered \cite{Blatt11}. 
Furthermore, in extending the setup  of \cite{polzik10} one could face up with the situation of using more than one 
laser beam (and related vacuum fluctuations), hence ending up with overlapping environments.

A similar situation can arise in cavity QED experiments with cavities 
hosting an ensembles of atoms and connected by fibres which play the role of an environment
\cite{cavityQED}.

Quite generally, when one has an array of atomic ensembles, 
like array optical traps loaded with neutral atoms \cite{otraps} or 
array of quantum dots \cite{qdots},
it may happen that the bath affecting one site can extend its effect over the neighbours sites.

%Other possible field of application is quantum dots coupled to leads. Electron transport between quantum dots trough %their coupling to leads can play the role of common qubits in our model \cite{dots}. Other example is a hybrid circuit-%QED consisting of double quantum dots in a common microwave cavity \cite{HQED} 

We are confident that the present study sheds further light on the dissipative quantum dynamics that is becoming increasingly exploited in quantum information processing. Specifically it should help in understanding how entanglement is induced by the interplay of environments that do not act separately.

%%%%%%%%%%%%%%%%%%%%%%%%%%%%%%%%%%%%%%%%%%%%%%%%%%%%%%%%%%
\acknowledgments
L. Memarzadeh would like to thank the University of Camerino for kind hospitality and INFN for financial support.
%%%%%%%%%%%%%%%%%%%%%%%%%%%%%%%%%%%%%%%%%%%%%%%%%%%%%%%%%%

\appendix

\section{Dynamics of the system when the initial excitation is in $C$}\label{AppA}

When the initial exitation is in $C$, 
the density matrix is given by \eqref{exprho}
and the dynamics is described by the
following set of differential equations for the time dependent coefficients:
\begin{align}
\label{ceqs}
\dot{c}_0&=4c_1+2N^2(c_2+c_3)+4(n-1)^2c_4\notag\\
&+4N(c_5+c_6)+8(n-1)c_7\notag\\
&+ 4N(n-1)(c_9+c_{10}),\cr\cr
\dot{c}_1&=-4c_1-2N(c_5+c_6)-4(n-1)c_7,\cr\cr
\dot{c}_2&=-2Nc_2-2c_5-2(n-1)c_9,\cr\cr
\dot{c}_3&=-2Nc_3-2c_6-2(n-1)c_{10},\cr\cr
\dot{c}_4&=-4(n-1)c_4-4c_7-2N(c_9+c_{10}),\cr\cr
\dot{c}_5&=-c_1-Nc_2-(N+2)c_5-(n-1)c_7\notag\\
&-Nc_8-2(n-1)c_9,\notag
\end{align}
\begin{align} 
\dot{c}_6&=-c_1-Nc_3-(N+2)c_6-(n-1)c_7\cr\cr
&-Nc_8-2(n-1)c_{10},\cr\cr
\dot{c}_7&=-2c_1-2(n-1)c_4-N(c_5+c_6)\notag\\
&-2nc_7-N(c_9+c_{10}),\cr\cr
\dot{c}_8&=-c_5-c_6-2Nc_8-(n-1)(c_9+c_{10}),\cr\cr
\dot{c}_9&=-Nc_2-(n-1)c_4-2c_5-c_7-Nc_8\notag\\
&-(N+2n-2)c_9,\cr\cr
\dot{c}_{10}&=-Nc_3-(n-1)c_4-2c_6-c_7-Nc_8\notag\\
&-(N+2n-2)c_{10}.
\end{align}
The initial conditions read $c_j(0)=\delta_{j,1}$ with $j=0,\ldots, 10$.

The solution of the set of differential equations \eqref{ceqs} can be  easily found.
First notice that due to symmetry between $A$ and $B$, it is $c_2=c_3 $, $c_5=c_6$,
$c_9=c_{10}$, furthermore it results
\begin{align}
\label{ceqssol}
c_0&= \dfrac{2  e^{-2(N+2n)t} }{(N+2n)} \left[e^{2(N+2n)t} -1\right],
\cr\cr
c_1&= \dfrac{e^{-2(N+2n)t}}{(N+2n)^2}  \left[(N+2n-2) e^{(N+2n)t}+2 \right]^2,
\cr\cr
c_2&=c_8= \dfrac{e^{-2(N+2n)t}}{(N+2n)^2} \left[e^{(N+2n)t}-1\right]^2,
\notag\\
c_4&=4c_2,
 \cr\cr
c_5&= \dfrac{e^{-2(N+2n)t}}{(N+2n)^2}\left[
-(N+2n-2)\ e^{2(N+2n)t} \right. \notag\\
&\left. \hspace{2.2cm} +(N+2n-4)\ e^{(N+2n)t} +2 \right],
\cr\cr
c_7&=2c_5,\quad c_8=c_2,\quad c_9=2c_2.
\end{align}

\section{Dynamics of the system when the initial excitation is in $A$}\label{AppB}

In the case that the initial excitation is in the ensemble $A$, the density matrix is given by \eqref{rhoexpprime}. 
Then, the dynamics is governed by the following differential equation for the time dependent coefficients:
\begin{align}\label{aeqs}
\dot{a}_0&=2a_1+2(N-1)^2a_2+2N^2a_3+4n^2 a_4 \notag\\
&+ 4(N-1)a_5+4n a_7 +4n(N-1)a_9+4Nn a_{10}, \cr\cr
\dot{a}_1&=-2a_1-2(N-1)a_5-2n a_7, \cr\cr
\dot{a}_2&=-2(N-1)a_2-2a_5-2n a_9, \cr\cr
\dot{a}_3&=-2N a_3-2n a_{10}, \cr\cr
\dot{a}_4&=-4n a_4-2a_7-2(N-1)a_9-2Na_{10}, \cr\cr
\dot{a}_5&=-a_1-(N-1)a_2-Na_5-n (a_7+a_9), \cr\cr
\dot{a}_6&=-(N+1)a_6-n a_7-(N-1)a_8-n a_{10}, \cr\cr
\dot{a}_7&=-a_1-n a_4-(N-1)a_5-N a_6+\notag\\
&-(2n+1)a_7-(N-1)a_9, \cr\cr
\dot{a}_8&=-a_6-(2N-1)a_8-n(a_9+a_{10}), \cr\cr
\dot{a}_9&=-(N-1)a_2-n a_4-a_5-a_7-Na_8+\notag\\
&-(N+2n-1)a_9, \cr\cr
\dot{a}_{10}&=-Na_3-n a_4-a_6-(N-1)a_8-(N+2n)a_{10}.\notag\\
\end{align}
The solutions, with initial conditions $ a_j(0)=\delta_{j,1}$ with $j=0,\ldots, 10$, read
\begin{align}
\label{aeqssol}
&a_0=\frac{e^{-2Nt}}{2N (N+2n)}\left[2(N+n)e^{2Nt}-(N+2n)-Ne^{-4 nt}\right], \cr\cr
&a_1=\frac{\left(N^2+2Nn-N-n\right) e^{-N t}}{N^2 (N+2 n)^2}\notag\\
   &\times\left[\left(N^2+2Nn-N-n\right)e^{Nt}+(N+2n)+Ne^{-2nt}\right]\notag\\
   &+\frac{e^{-2 N t}}{4N^2(N+2n)^2}\left[(N+2n)+Ne^{-2nt}\right]^2, \cr\cr
   &a_2=\frac{e^{-2 N t}  }{4 N^2 (N+2 n)^2} \left[N+2n+Ne^{-2 n t}-2 e^{N t}(N+n)\right]^2, \cr\cr  
&a_3=\frac{e^{-2 N t}}{4 N^2(N+2 n)^2} \left[-2 n e^{N t}+(N+2n)-Ne^{-2 n t}\right]^2, \cr\cr
&a_4=\frac{ e^{-2 (N+2 n)t}}{(N+2 n)^2} \left[e^{(N+2 n)t}-1\right]^2,\cr\cr
&a_5=\frac{\left(N^2+2 N n-2N-2 n\right) e^{-N t}}{2N^2 (N+2 n)^2}\notag\\
   &\times\left[-2(N+n)e^{Nt}+(N+2n)+Ne^{-2nt}\right]\notag\\
   &+\frac{e^{-2 N t}}{4N^2(N+2n)^2}\left[(N+2n)+Ne^{-2nt}\right]^2, \notag
\end{align}
\begin{align}   
&a_6=\frac{e^{-Nt}}{4N^2(N+2n)^2} \left[4n\left(N^2+2Nn-N-n\right)e^{Nt} \right.\notag\\
   &\left.-2(N+2n)^2(N-1)-(N+2n)^2e^{-N t}\right.\notag\\
   &\left.+2N^2\left( N+2n-1\right)e^{-2nt}+N^2e^{-(N+4n)t}\right], \cr\cr
&a_7=\frac{e^{-Nt}}{2 N (N+2 n)^2}\left[2 \left(n- N^2-2 N n+N\right)e^{Nt}\right.\notag\\
      &\left.-(N+2 n) +\left(2 N^2-2n+4Nn-3N\right) e^{-2 nt} \right.\notag\\
      &\left.  +(N+2 n) e^{-(N+2n)t} +Ne^{-(N+4 n)t} \right], \cr\cr
&a_8=\frac{e^{-2 Nt}}{4 N^2 (N+2 n)^2}
      \left[-4n(N+n)e^{2Nt}+
      2(N+2n)^2 e^{Nt}  \right.\notag\\
      &\left.-(N^2+2n)^2-2 N^2e^{(N-2n)t}-N^2e^{-4nt}\right], \cr\cr
&a_9=\frac{\left[e^{(N+2n)t}-1\right] e^{-2 (N+n) t} }{2 N (N+2 n)^2}\notag\\
 &\times [2 (N+n) e^{N t} -(N+2n)-N e^{-2nt}], \cr\cr
&a_{10}=\frac{ \left[e^{(N+2 n)t}-1\right] e^{-2 (N+ n)t}}{2 N (N+2 n)^2}\notag\\
 &\times  \left[-2 n e^{N t}+(N+2n) -N e^{-2 n t}\right].\notag\\
\end{align}

%%%%%%%%%%%%%%%%%%%%%%%%%%%%%%%%%%%%%%%%%%%%%%%%%%%%%%%%%%%%%%%%%%%%%%%%%%%%%%%%%%%%%%%%%%%%%%%%%%%%%%%%%%%%%%%%%%%%%%%%%%%%%%%%%%%%%%%%%%%%%%%%%%%%%%%%%%%%%%%%%%%%%%%%%%%

\end{document}